\documentclass{WileyMSP-template}

\usepackage[T1]{fontenc}
\usepackage[latin9]{luainputenc}
\usepackage{color}
\usepackage{pifont}
\usepackage{textcomp}
\usepackage{pmboxdraw}
\usepackage{amstext}
\usepackage{amssymb}
\usepackage{graphicx}
\usepackage{esint}
\usepackage{subscript}
\usepackage{tabularx}
\usepackage{float}
\usepackage[export]{adjustbox}
\usepackage{hyperref}
\usepackage{upgreek,textgreek}
\usepackage{siunitx}
\usepackage{babel}
\usepackage{hyperref}
\usepackage{ragged2e}
\usepackage{caption}
\usepackage{cite} 

\begin{document}
\begin{sloppypar}
\pagestyle{fancy}
\rhead{\includegraphics[width=2.5cm]{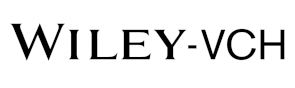}}

\captionsetup[figure]{labelfont={bf},labelformat={default},labelsep=period,name={Figure}}

\begin{center}
\title{Spin-phonon scattering-induced low thermal conductivity in a van der Waals layered ferromagnet Cr$_2$Si$_2$Te$_6$}
\end{center}
\maketitle



\author{Kunya Yang}
\author{Hong Wu}
\author{Zefang Li}
\author{Chen Ran}
\author{Xiao Wang}
\author{Fengfeng Zhu}
\author{Xiangnan Gong}
\author{Yan Liu}
\author{Guiwen Wang}
\author{Long Zhang}
\author{Xinrun Mi}
\author{Aifeng Wang}
\author{Yisheng Chai}
\author{Yixi Su}
\author{Wenhong Wang}
\author{Mingquan He \textcolor{blue}{$^*$}}
\author{Xiaolong Yang \textcolor{blue}{$^*$}}
\author{Xiaoyuan Zhou \textcolor{blue}{$^*$}}
\date{\today}


\begin{affiliations}
Kunya Yang, Chen Ran, Xiangnan Gong, Yan Liu, Guiwen Wang, Long Zhang, Xinrun Mi, Yisheng Chai, Mingquan He, Xiaolong Yang, Xiaoyuan Zhou\\
College of Physics \& Center of Quantum Materials and Devices \& Analytical and Testing Center\\
Chongqing University\\
Chongqing 401331, China\\
E-mail:mingquan.he@cqu.edu.cn\\
E-mail:yangxl@cqu.edu.cn\\
E-mail:xiaoyuan2013@cqu.edu.cn\\

Hong Wu\\
School of Science, Chongqing University of Posts and Telecommunications\\  Chongqing 400065, China\\

Zefang Li\\
Ultrafast Electron Microscopy Laboratory, The MOE Key Laboratory of Weak-Light Nonlinear Photonics,School of Physics\\
Nankai University\\
Tianjin 300071, China\\
Beijing National Laboratory for Condensed Matter Physics
Institute of Physics\\
Chinese Academy of Sciences\\
Beijing 100190, China\\

Xiao Wang,Yixi Su\\
J\"ulich Centre for Neutron Science (JCNS) at Heinz Maier-Leibnitz Zentrum (MLZ), Forschungszentrum J\"ulich GmbH, Lichtenbergstr. 1, D-85747 Garching, Germany\\

Fengfeng Zhu\\
State Key Laboratory of Functional Materials for Informatics\\
Shanghai Institute of Microsystem and Information Technology\\
Chinese Academy of Sciences, Shanghai 200050, China\\

Wenhong Wang\\
Beijing National Laboratory for Condensed Matter Physics\\
Institute of Physics\\
Chinese Academy of Sciences, Beijing 100190, China\\
Tiangong University\\
Tianjin, 300387, China\\
\end{affiliations}



\justifying
\begin{abstract}
Layered van der Waals (vdW) magnets are prominent playgrounds for developing magnetoelectric, magneto-optic and spintronic devices. In spintronics, particularly in spincaloritronic applications, low thermal conductivity ($\kappa$) is highly desired.  Here, by combining thermal transport measurements with density functional theory calculations, we demonstrate low $\kappa$ down to 1 W m$^{-1}$ K$^{-1}$ in a typical vdW ferromagnet Cr$_2$Si$_2$Te$_6$. In the paramagnetic state, development of magnetic fluctuations way above $T_\mathrm{c}=$ 33 K strongly reduces $\kappa$ via spin-phonon scattering, leading to low $\kappa \sim$ 1 W m$^{-1}$ K$^{-1}$ over a wide temperature range, in comparable to that of amorphous silica. In the magnetically ordered state, emergence of resonant magnon-phonon scattering limits $\kappa$ below $\sim$ 2 W m$^{-1}$ K$^{-1}$, which would be three times larger if magnetic scatterings were absent. Application of magnetic fields strongly suppresses the spin-phonon scattering, giving rise to large enhancements of $\kappa$. Our calculations well capture these complex behaviours of $\kappa$ by taking the temperature- and magnetic-field-dependent spin-phonon scattering into account. Realization of low $\kappa$ which is easily tunable by magnetic fields in Cr$_2$Si$_2$Te$_6$, may further promote spincaloritronic applications of vdW magnets. Our theoretical approach may also provide a generic understanding of spin-phonon scattering, which appears to play important roles in various systems.
\end{abstract}


\section{Introduction}
Discoveries of intrinsic two-dimensional (2D) magnetism in van der Waals (vdW) magnets, such as Cr$_2$(Si, Ge)$_2$Te$_6$, Cr(Cl, Br, I)$_3$ and Fe$_3$GeTe$_2$, open up new avenues for exploring novel magnetic sates and device applications \cite{Gong2017CGT,Huang2017CI,Fei2018FGT}. In real device applications, various optimized physical aspects are desired, such as electrical, optical and thermal properties, in addition to novel magnetism. For example, in spincaloritronic devices utilizing the coupling of spin, charge and entropy flow, low thermal conductivity ($\kappa$) is highly favored \cite{Bauer2012SC,Kirihara2012SS}. Despite intensive studies, little attention has been paid to thermal properties of vdW magnets. Notably, sizable spin Seebeck effect has been experimentally observed in Cr$_2$(Si, Ge)$_2$Te$_6$ \cite{Ito_CST_SS}. Giant spin Seebeck effect is also theoretically expected in CrI$_3$ \cite{Marfoua_CrI3_SS,Tan_CrI3_SS}. It is of great importance to realize low $\kappa$ in these vdW magnets, which could further facilitate device implementations. 

In the phonon gas picture, heat conduction in crystalline solids can be treated as propagation and scattering of phonon wave packets, resembling moving gas particles (particle-like phonons) \cite{Qian2021}.  Intrinsically low thermal conductivity is typically found in crystals possessing strong anharmonicity or complex structures, in which the phonon mean free path (MFP) is significantly reduced, as found in SnSe and Tl$_3$VSe$_4$, cage structured clathrates and skutterudites \cite{Zhao2014,Mukhopadhyay2018,Sales1996,Christensen2008,Qian2021}. Notably, when the phonon MFP approaches the interatomic separation, wavevectors of phonons are no longer well defined. In this case, thermal transport is accomplished in a diffusive fashion (diffuson-like phonons), leading to glass-like low thermal conductivity < 1 W m$^{-1}$ K$^{-1}$ \cite{Anderson1972,Allen1999}. Although typically found in highly disordered systems, glass-like heat transport can appear in single crystals containing complex structures \cite{Cohn1999,Wybourne1984,Hermann2006,Xia2020,Xiayi2020} or atomic tunneling effects \cite{Sun2020}. 

In vdW magnets, the 2D magnetism is naturally associated with strong spin fluctuations, which can strongly suppress $\kappa$ via spin-phonon scattering. This offers another prominent channel to realize intrinsic low $\kappa$ without invoking strong anharmonicity, complex structures or disorders.  Indeed, significantly suppressed $\kappa$ driven by spin-phonon scattering has been observed in various magnetic systems, such as conventional 2D magnets K$_2$V$_3$O$_8$, Nd$_2$CuO$_4$ and CrCl$_3$ \cite{Jin2003,Sales2002,Pocs2020}, antiferromagnetic (AFM) ferroelectrics $X$MnO$_3$ ($X$ = Y, Ho, Lu, Sc, Ca) \cite{Sharma2004,Chiorescu2008,Wangx2010}, metallic spin ice Pr$_2$Ir$_2$O$_7$ \cite{Uehara2022}, Kitaev quantum spin liquid candidates $\alpha$-RuCl$_3$ and Na$_2$Co$_2$TeO$_6$ \cite{Hirobe_Rucl3,Leahy2017,Hentrich2018,Xiaochen2021,Yang-NCTO_therHall}, and nuclear energy fuel UO$_2$ \cite{Gofryk2014UO2}. Still, low $\kappa$ down to $\sim$ 1 W m$^{-1}$ K$^{-1}$ induced by spin-phonon scattering is rarely found. Moreover, although spin-phonon scattering has been discussed in various circumstances, a fundamental understanding of temperature- and
magnetic-field-dependent spin-phonon scattering effects is not well established.  
           
Here, we present spin-phonon scattering-induced low $\kappa$ in single crystals of a typical vdW ferromagnet (FM) Cr$_2$Si$_2$Te$_6$ (CST). Its 2D magnetic nature enables strong magnetic fluctuations surviving up to $\sim$ 200 K,  giving rise to low $\kappa \sim$ 1 W m$^{-1}$ K$^{-1}$ at all temperatures above 
 $T_\mathrm{c}$ = 33 K. Large positive thermal magnetoconductivity is found below about 100 K, allowing easy manipulation of $\kappa$ by external magnetic fields. Our theoretical calculations can well reproduce these magnetic scattering effects, which may provide a generic fundamental understanding of spin-phonon scattering. The observed low thermal conductivity and its easy tunability in CST may further advance device implementations of vdW magnets, particularly in spincaloritronic applications.

\section{Results}
\subsection{Structural and thermodynamic properties}
\begin{figure*}[t]
\includegraphics[scale=0.5,center]{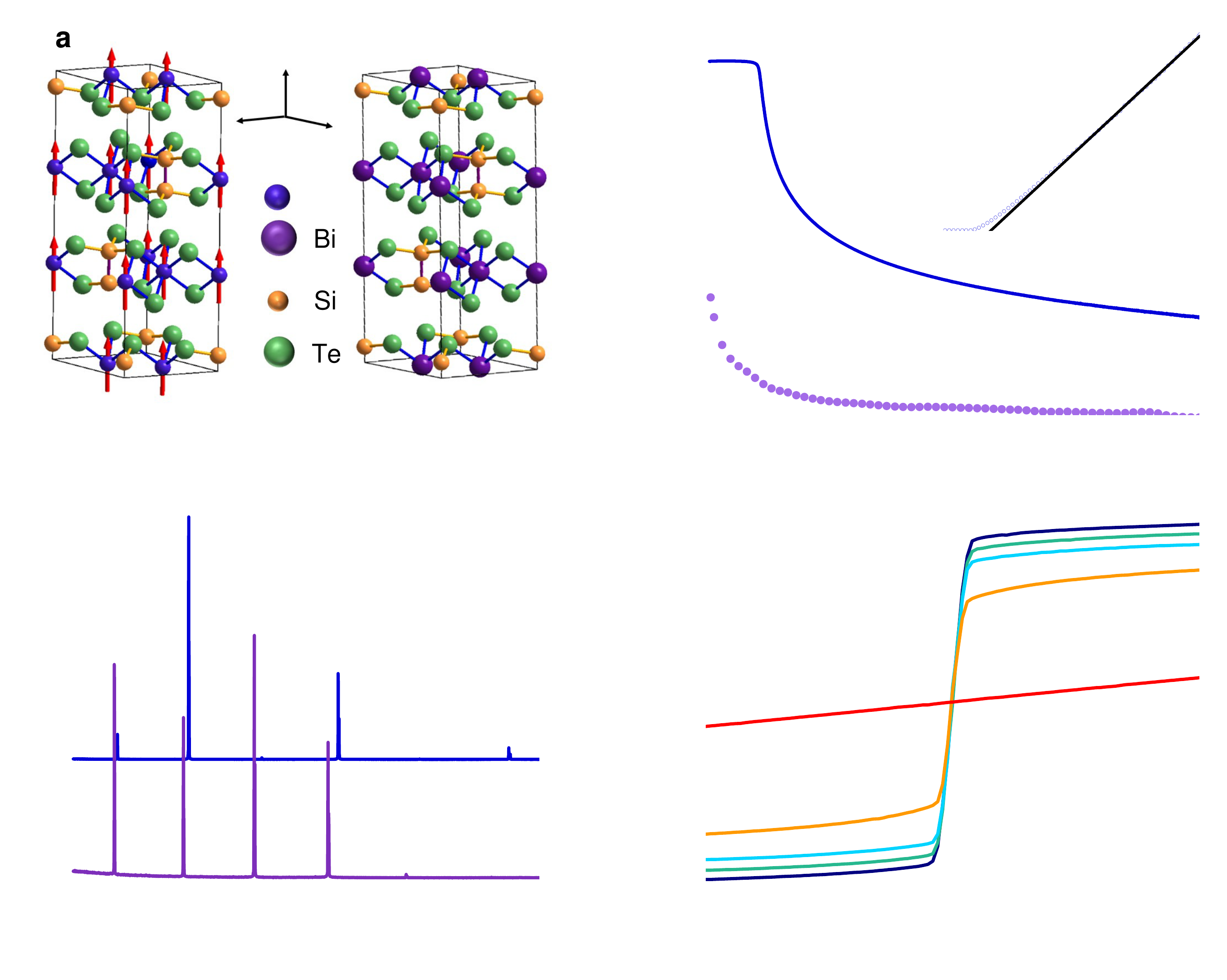}
\caption{\textbf{Structural and magnetic properties of Cr$_2$Si$_2$Te$_6$ (CST) and Bi$_2$Si$_2$Te$_6$ (BST).} a) Crystal structures of CST and its non-magnetic sister compound BST. Arrows on Cr atoms represent magnetic moments, which order ferromagnetically along the $c$-axis below $T_\mathrm{c}=33$ K. b) Single crystal X-ray diffraction data of CST and BST. c) Temperature-dependent magnetization measured in zero-field-cooled (ZFC) with an magnetic field $B=0.1$ T applied along the $c$-axis. The black arrow marks the FM transition of CST at $T_\mathrm{c}=33$ K. The inset in (c) shows the inverse susceptibility together with a Curie-Weiss (CW) fitting (black sold line). d) Isothermal magnetization of CST measured as a function of magnetic field.   
\label{fig:1}}
\end{figure*}

CST is a van der Waals layered material crystallizing in a rhombohedral structure ($R\overline{3}$) \cite{OUVRARD198827,Casto2015}, as shown in Figure \ref{fig:1}a. Each layer contains edge-sharing Cr-Te octahedra, and layers are stacked along the $c$-axis in an $ABC$ fashion. The Cr$^{3+}$ ($S=3/2$) ions form a honeycomb lattice within the $ab$ plane, and order ferromagnetically along the $c$-axis below $T_\mathrm{c} = 33$ K \cite{Carteaux_1995,Williams2015,Zhangjiaxin2019}. For comparison, an iso-structural non-magnetic sister compound Bi$_2$Si$_2$Te$_6$ (BST), which has not been experimentally reported before, is also studied here. Replacement of Cr by Bi with larger atomic sizes leads to moderate lattice expansion along both the $a$-axis and the $c$-axis, as seen in single crystal X-ray diffraction data displayed in  Figure \ref{fig:1}b. The measured lattice parameters of BST agree well with an earlier theoretical study \cite{Liping2019}. 

The magnetic properties of typical crystals of CST and BST are presented in Figures \ref{fig:1}c, d. For BST, a paramagnetic behavior with vanishing magnetic moments is seen in the temperature dependence of magnetization all the way from room temperature down to 2 K, implying its non-magnetic nature. For CST, a clear long-range FM transition is found at $T_\mathrm{c}=33$ K, agreeing well with other studies \cite{Carteaux_1995,Williams2015,Zhangjiaxin2019,Lizefang2021,Zhufengfeng2021}. A Curie-Weiss (CW) fit can well describe the inverse susceptibility $1/\chi$ of CST above 150 K, as presented in the inset of Figure \ref{fig:1}c.  At lower temperatures, $1/\chi$ deviates from the linear Curie-Weiss expectation, suggesting the appearance of sizable short-range spin fluctuations \cite{Casto2015,Williams2015,Milosavljevi2018,Ron2019,Lizefang2021}. The short-range magnetic fluctuations even survives up to room temperature as seen by inelastic neutron scattering experiments \cite{Williams2015}. Normally, magnetic fluctuations only appear in the vicinity of $T_\mathrm{c}$ in typical magnetic systems. The exceptionally large temperature window spanned by spin fluctuations in CST is driven by its 2D magnetic nature, which supports a large Ginzburg number \cite{Lizefang2021}. These short-range spin fluctuations turn out to be detrimental for phonon heat transport due to spin-phonon scattering, as we will see shortly. The isothermal magnetization is displayed in Figure \ref{fig:1}d. Little hysteresis is found and the coercive field is practically zero.  Moderate fields ($\sim$ 0.2 T) applied out-of-plane can polarize the magnetic moments in the FM ordered state. The saturation moment only reaches $\sim$ 2.8 $\mu_\mathrm{B}$ at 2 K,  which is smaller than the theoretical value of 3.87 $\mu_\mathrm{B}$ for Cr$^{3+}$. This again implies considerable contributions from magnetic fluctuations. 

\begin{figure*}[t]
\includegraphics[scale=0.5,center]{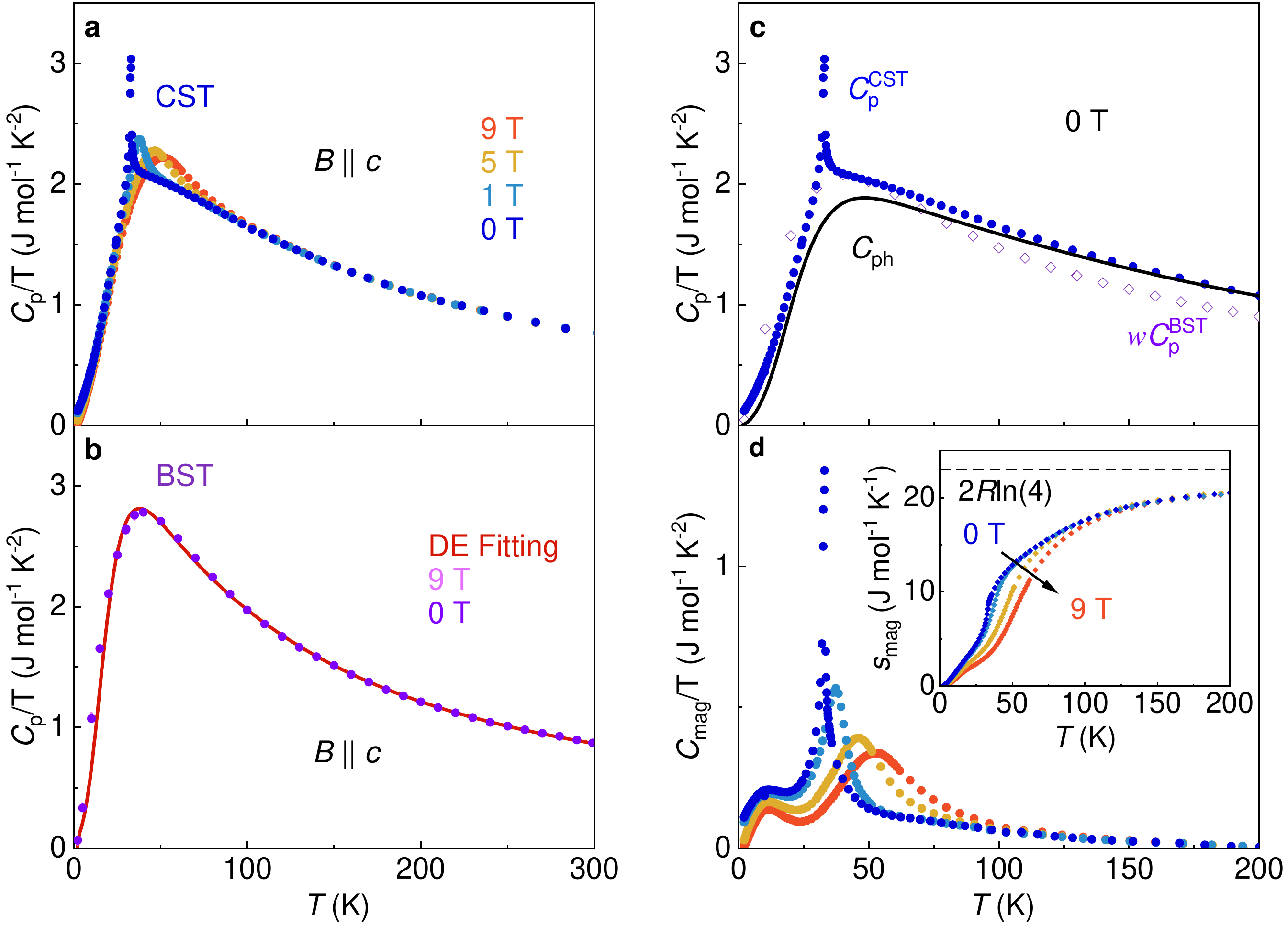}
\caption{\textbf{Specific heat of CST and BST measured in various magnetic fields applied out-of-plane.} a,b)  Solid line in (b) is a theoretical fitting according to the Debye-Einstein scenario. c) Comparison of the specific heat of CST ($C_\mathrm{p}^\mathrm{CST}$) to that of BST ($C_\mathrm{p}^\mathrm{BST}$), and to the phonon background ($C_\mathrm{ph}$) evaluated from DFT calculations. The data of BST is scaled by a constant $w = 0.75$, which is the molar mass scaling factor between CST and BST. d) The magnetic specific heat $C_\mathrm{mag}$ of CST obtained by subtracting the phonon contributions $C_\mathrm{ph}$ from the total $C_\mathrm{p}^\mathrm{CST}$. The inset in (d) presents the magnetic entropy $s_\mathrm{mag}$ evaluated by integrating the $C_\mathrm{mag}$ displayed in the main panel of (d). The horizontal dash line in the inset of (d) labels the theoretical value of 2$R$ln(4) ($R$ is the ideal-gas constant) for an $S=3/2$ spin configuration with 2 Cr atoms per unit cell.}
\label{fig:2}
\end{figure*}

Figure \ref{fig:2} presents the specific heat $C_\mathrm{p}$ of CST and BST. In zero magnetic field, the second order FM transition in CST is evidenced by a $\lambda$-shaped peak at $T_\mathrm{c}=33$ K, as shown in Figure \ref{fig:2}a. The FM transition is smeared out gradually in finite magnetic fields, showing a broad bump which shifts to higher temperatures. The magnetic field response of $C_\mathrm{p}$ becomes negligible above  $\sim$ 150 K, which also suggests that magnetic fluctuations persist up to at least  150 K. On the other hand, $C_\mathrm{p}$ measured in 0 and 9 T are identical in BST, which can be fitted nicely using the phononic Debye-Einstein model (see solid line in Figure \ref{fig:2}b), as expected for a non-magnetic material. To quantitatively evaluate the magnetic contribution $C_\mathrm{mag}$ in CST, we attempted to use BST as the phonon background $C_\mathrm{ph}$. However, as shown in Figure \ref{fig:2}c, this approach does not work well. The scaled specific heat of BST, $wC^\mathrm{BST}_\mathrm{p}$, deviates substantially from that of CST ($C^\mathrm{CST}_\mathrm{p}$), especially at high temperatures. Here, $w = 0.75$ is the molecular mass ratio between CST and BST. This discrepancy likely originates from the large differences in lattice parameters between CST and BST. A better approximation of $C_\mathrm{ph}$ is found from the phonon dispersion obtained by density functional theory (DFT) calculations (black solid line in Figure \ref{fig:2}c). The magnetic contributions are thus obtained for CST as $C_\mathrm{mag} = C_\mathrm{p}-C_\mathrm{ph}$, and the results are presented in Figure \ref{fig:2}d. It is seen that $C_\mathrm{mag}$ appears well above $T_\mathrm{c}$, extending up to about 200 K. As shown in the inset of Figure \ref{fig:2}d, by integrating $C_\mathrm{mag}$ out, the obtained magnetic entropy $s_\mathrm{mag}$ nearly approaches the theoretical value 2$R$ln(4) ($R$ is the ideal-gas constant) above 200 K for various magnetic fields. This implies that the calculated phonon dispersion and $C_\mathrm{ph}$ are in close proximity to the phonon contributions of CST, and that short-range spin correlations survive up to about 200 K. 

\begin{figure*}[t]
\includegraphics[scale=0.5,center]{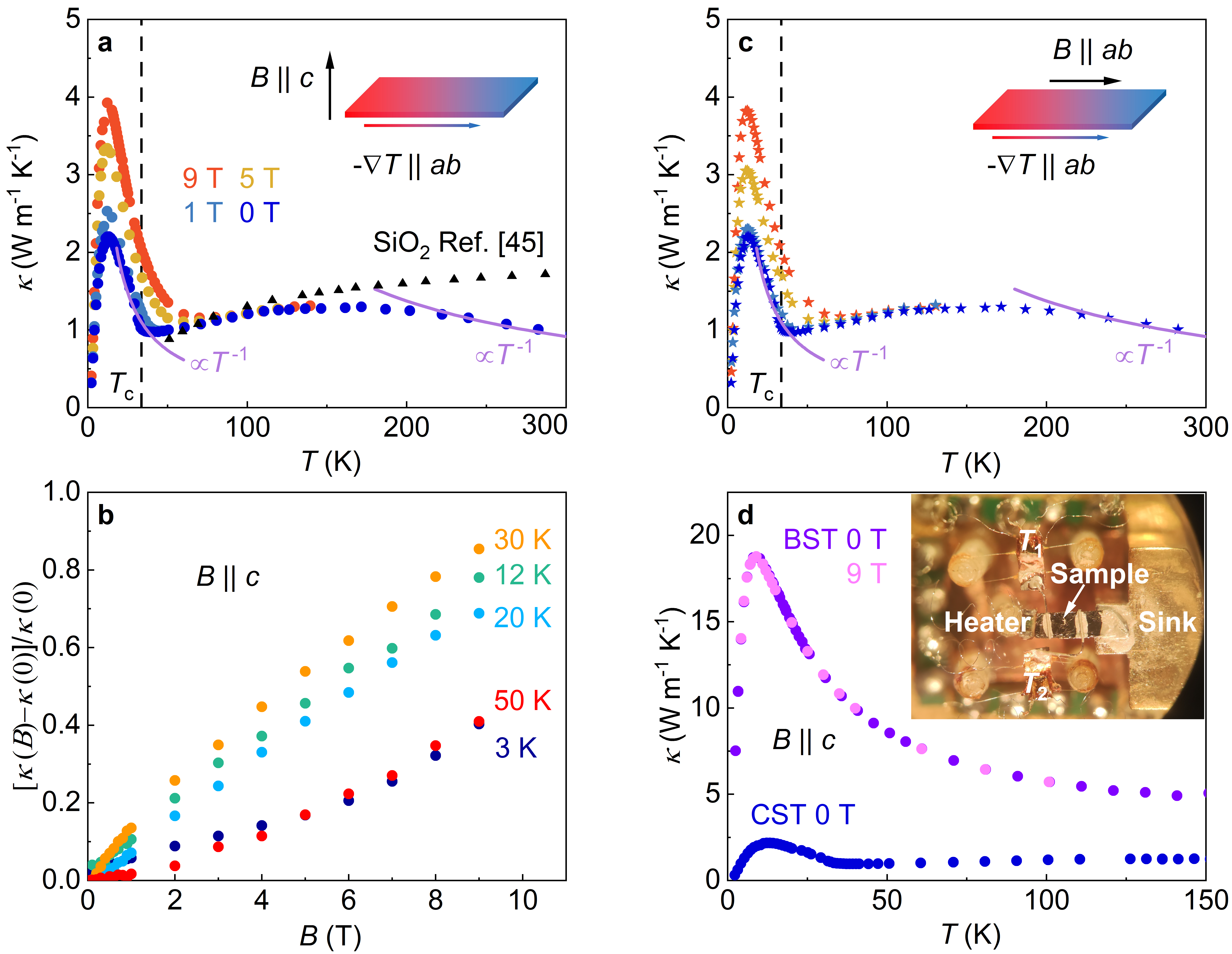}
\caption{\textbf{Thermal conductivity of CST and BST.} a,c)Temperature dependence of $\kappa$ for CST measured in various magnetic fields applied along the $c$-axis and $ab$ plane. The thermal gradient $-\nabla T$ was sent within the $ab$ plane (see insets). Solid purple lines simulate the $T^{-1}$ temperature dependence. The $\kappa$ of amorphous silica is also plotted for comparison (black triangles Ref. \cite{cahill2010low}). b) Thermal magnetoconductivity recorded at selective temperatures for $B\parallel c$. d) Comparison of $\kappa$ in CST and BST. \textcolor{black}{The inset in d) shows a photograph of the set-up used in thermal transport measurements.}}
\label{fig:3}
\end{figure*}

\subsection{Thermal conductivity}
In Figure \ref{fig:3}, we present the measured in-plane thermal conductivity of CST and BST.  In zero magnetic field, above $\sim$ 200 K, $\kappa$ of CST increases with cooling, following a typical $ T^{-1}$ law (see solid purple lines in Figures \ref{fig:3}a,c), which implies a normal phonon transport behavior governed by Umklapp scattering \cite{Lichunhua2018,Yangxiaolong2021}. The observed low thermal conductivity of $\sim$ 1 W K$^{-1}$ m$^{-1}$ and the $ T^{-1}$ temperature dependence near 300 K agree quantitatively with those measured within 300 - 500 K using a laser flash method  \cite{Lefevre2017}.   Intriguingly, upon further cooling, $\kappa$ decreases monotonically with cooling rather than obeying the $T^{-1}$ trend, and low thermal conductivity $\sim$ 1 W K$^{-1}$ m$^{-1}$ is realized all the way down to $T_\mathrm{c}$. Similar behavior has been reported earlier \cite{Casto2015}, and  also appears in a sister compound Cr$_2$Ge$_2$Te$_6$ \cite{Liuyu2022}. The small values and peculiar temperature dependence of $\kappa$ are  similar to those in amorphous silica (see black triangles in Figure \ref{fig:3}a), likely suggesting glass-like thermal transport \cite{Cohn1999,Wybourne1984,Hermann2006,Xia2020,Xiayi2020,Casto2015,Liuyu2022,Sun2020}.  However, as we will show in Figure \ref{fig:4},  diffuson-like heat conduction only contributes moderately, and spin-phonon scattering  becomes dominant when magnetic fluctuations appear below 200 K.  Further cooling below $T_\mathrm{c}$, $\kappa$ seemingly recovers to  typical phonon thermal transport with the familiar $T^{-1}$ temperature dependence. A characteristic peak is found around 12 K with $\kappa\sim$ 2 W K$^{-1}$ m$^{-1}$, which is the maximum value observed within the entire temperature range studied here. 

In finite magnetic fields, $\kappa$ of CST is significantly enhanced, and similar results are found for both out-of-plane ($B\parallel c$, Figure \ref{fig:3}a) and in-plane ($B\parallel ab$, Figure \ref{fig:3}c) configurations. The $\kappa$ is nearly doubled in 9 T around 12 K. This positive thermal magnetoconductivity persists in the paramagnetic state up to $\sim$ 100 K. As shown in Figure \ref{fig:3}b, $\kappa(B)$ increases monotonically with increasing magnetic field both above and below $T_\mathrm{c}$, and no saturation is found up to 9 T. Magnetic moments are well saturated in 9 T below $T_\mathrm{c}$ (see Figure \ref{fig:1}), and a single magnetic domain is expected. The non-saturation $\kappa(B)$ up to 9 T is thus not related to phonon scattering by magnetic domains. Note that CST is highly insulating, and the electronic contributions $\kappa_\mathrm{el}$ are negligible (see Supplemental Material \cite{supplemental}). Thus, the field-induced enhancement of $\kappa$ has phononic origins.  The observed large thermal magnetoconductivity apparently is not originated from conventional scatterings of phonons, including anharmonicity, impurities, point defects, and sample boundaries, which are independent of magnetic field. The remaining possibility goes to spin degrees of freedom that are highly field sensitive, i.e., spin-phonon scattering, as found in various magnetic systems \cite{Jin2003,Sales2002,Pocs2020,Sharma2004,Chiorescu2008,Leahy2017,Hentrich2018,Xiaochen2021,LiQ2013,Wangx2010,Uehara2022}.  This scenario is further supported by comparing the thermal conductivity of CST and BST (see Figure \ref{fig:3}d). Clearly, $\kappa$ of non-magnetic BST shows a typical phononic heat transport, which remains intact in the presence of magnetic fields. Compared to CST, the $\kappa$ of BST is one order of magnitude larger. We note that similar effects appear in a cubic antiferromagnet UO$_2$, which is the benchmark fuel for nuclear power generation. Compared to a non-magnetic sister material ThO$_2$, the thermal conductivity of UO$_2$ is reduced by a factor up to about 50 due to spin-phonon scattering effects \cite{Gofryk2014UO2}. However, the low thermal conductivity of UO$_2$ puts challenges in reactor design and safety.  Understanding of the spin-phonon scattering processes can also provide crucial insights to such research fields where magnetic scattering-induced low thermal conductivity is troublesome.

\subsection{Spin-phonon scattering}
\begin{figure*}[t]
\includegraphics[scale=0.4,center]{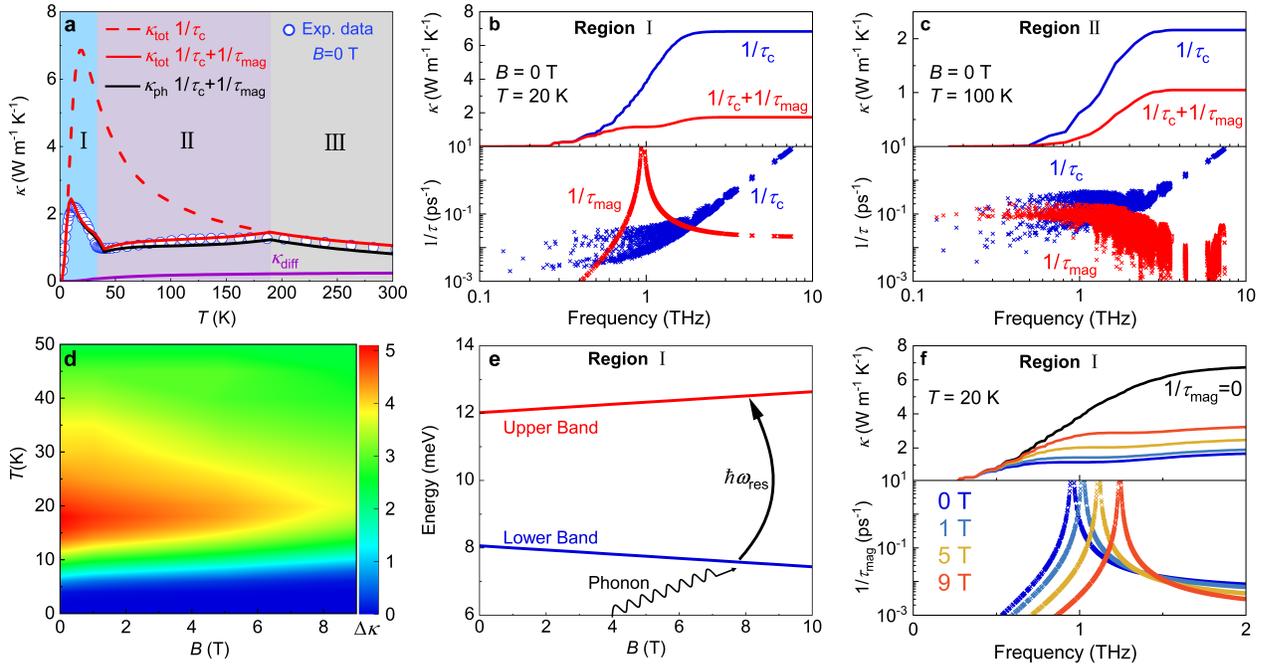}
\caption{\textbf{Calculated thermal conductivity of CST.} a) By considering conventional ($1/\tau_\mathrm{c}$)  and magnetic scattering ($1/\tau_\mathrm{mag}$) rates, the calculated total thermal conductivity $\kappa_\mathrm{tot}=\kappa_\mathrm{ph}+\kappa_\mathrm{diff}$ (red solid line) follows experimental data (blue circles) nicely in all temperature regions. Black and purple curves are contributions from propagating phonons ($\kappa_\mathrm{ph}$) and diffuson-like ($\kappa_\mathrm{diff}$) channels, respectively. Calculated $\kappa_\mathrm{tot}(1/\tau_\mathrm{c})$ without involving magnetic scattering is plotted for comparison (red dash line). b) and c) Comparison of  $1/\tau_\mathrm{c}$ and $1/\tau_\mathrm{mag}$, along with cumulative thermal conductivity evaluated at 20 K, and 100 K, respectively. d) Temperature- and field-dependent differential thermal conductivity $\Delta\kappa=\kappa_\mathrm{tot}(1/\tau_\mathrm{c})-\kappa_\mathrm{tot}(1/\tau_\mathrm{c}+1/\tau_\mathrm{mag})$.  e) Schematic illustration of resonant magnon-phonon scattering at a field-dependent resonant frequency $\omega_\mathrm{res}$ in the FM ordered state. f) The field dependence of  $1/\tau_\mathrm{mag}$ and corresponding cumulative thermal conductivity calculated at 20 K.}
\label{fig:4}
\end{figure*}

To quantitatively evaluate the effects of magnetic scatterings of phonons on heat transport, we have performed first-principles calculations. A unified theory of lattice heat transport considering both contributions from particle-like propagation ($\kappa_\mathrm{ph}$) and diffuson-like phonons ($\kappa_\mathrm{diff}$) was used \cite{Simoncelli2019}. Conventional phonon scattering events arising from anharmonicity, isotope, boundaries, and impurities, are described in the conventional phonon scattering rate $1/\tau_\mathrm{c}$. More importantly, to account for the observed temperature- and magnetic-field-dependent $\kappa$, additional contributions from magnetic scattering, $1/\tau_\mathrm{mag}$, are also considered below $\sim$ 200 K. More computational details and exact forms of $1/\tau_\mathrm{mag}$ can be found below and in Supplemental Material \cite{supplemental}.  The calculated zero-field thermal conductivity is shown in Figure~\ref{fig:4}a. To highlight the distinct behaviors and mechanisms of $\kappa$ in different temperature intervals, we split the temperature window into three regions: FM ordered state ($T\le T_\mathrm{c}$, region I), short-range spin fluctuation state ($T_\mathrm{c}<T<190$ K, region II), conventional paramagnetic phase ($T\ge 190$ K, region III).  It is seen that the calculated total thermal conductivity (red solid line), $\kappa_\mathrm{tot}=\kappa_\mathrm{ph}+\kappa_\mathrm{diff}$,  well describes the experimental data across the entire temperature range. Evidently, diffuson-like phonons only contribute moderately, and propagating phonons dominate in heat conduction. Thus, the unusual temperature dependence of $\kappa$  found in region II is unlikely originated from glass-like thermal transport. Notably, once the magnetic scattering rate is switched off, the calculated total $\kappa_\mathrm{tot}$ (red dash line in Figure~\ref{fig:4}a) recovers to the familiar phononic transport with much larger values in regions I and II, compared to $\kappa_\mathrm{tot}$ calculated invoking finite $1/\tau_\mathrm{mag}$. These results highlight the significance of $1/\tau_\mathrm{mag}$ in determining the magnitudes and temperature dependence of $\kappa_\mathrm{tot}$ when spin degrees of freedom become important in regions I and II. In region III, spin fluctuations become subdominant, and $\kappa$ is restored to the typical Umklapp scattering-dominated phonon transport.

To gain further insights into the importance of spin-phonon scattering in regions I and II, Figures~\ref{fig:4}b,c display the calculated scattering rates, along with the corresponding cumulative thermal conductivity. 
\textcolor{black}{In the FM ordered state (region I), the magnon-phonon scattering can happen in a resonant fashion in which a phonon having a specific energy of $\hbar\omega_\mathrm{res}$ excites a magnon from a lower band to an upper band with an energy difference of $\hbar\omega_\mathrm{res}$ \cite{Hofmann2001,Jin2003,Sales2002,Leahy2017,Hentrich2018,Xiaochen2021} (see also Figure~\ref{fig:4}e). The resonant magnon-phonon scattering rate can be modeled in a phenomenological picture \cite{Sales2002,Yang2022,Hentrich2018,Prasai2018}:}   
\begin{equation}
1/\tau_\mathrm{mag}(\omega)=M_\mathrm{res}\frac{\omega^{4}}{(\omega^{2}-\omega_\mathrm{res}^{2})^{2}}\frac{\mathrm{exp}(-\frac{\hbar\omega_\mathrm{res}}{k_{B}T})}{1+\mathrm{exp}(-\frac{\hbar\omega_\mathrm{res}}{k_{B}T})},
\label{eq:magnetic_scattering}
\end{equation}
\textcolor{black}{where $\omega$ is the phonon frequency, $M_\mathrm{res}$ denotes the magnon-phonon coupling constant, and $\frac{\omega^4}{(\omega^2-\omega_\mathrm{res}^2)^2}$ stands for the resonant scattering cross section at the resonant frequency $\omega_\mathrm{res}$. The Boltzmann temperature-dependent part describes the thermal population of resonant magnon-phonon modes. Note that this  phenomenological model is applicable in magnon bands with well-defined energy-momentum dispersion, which is indeed the case for CST as seen by neutron scattering experiments \cite{Williams2015,Zhufengfeng2021}.}
As shown in Figure~\ref{fig:4}b, the resonant magnon-phonon scattering gives a diverging magnetic scattering rate $1/\tau_\mathrm{mag}$ at a resonant frequency $\omega_\mathrm{res}$ around 1 THz (see the bottom panel in Figure~\ref{fig:4}b). As presented in the upper panel of Figure~\ref{fig:4}b, the cumulative thermal conductivity is dominated by phonons within the frequency window of 0 $\sim$ 2 THz. In this region, $1/\tau_\mathrm{mag}$ is mostly comparable to or even much stronger than $1/\tau_\mathrm{c}$. Compared with phonon transport involving $1/\tau_\mathrm{c}$ only, the thermal conductivity is reduced by a factor of $\sim$ 3.5 in the presence of magnon-phonon scattering.  

In region II, the long-range magnetic order is destroyed, leaving short-range spin fluctuations which can scatter phonons via spin-phonon coupling \cite{Sharma2004,Chiorescu2008,Bansal207,Hentrich2018,Wangx2010,Uehara2022,LiQ2013}. In a phenomenological picture, short-range spin fluctuations can be viewed as spherical scatterers with an average cross section of $\pi(\xi/2)^2$, where $\xi$ is the spin correlation length. According to the kinetic theory, the spin-phonon scattering rate can then be approximated as \cite{Sharma2004,Chiorescu2008,Bansal207}, 
\begin{equation}
 1/\tau_\mathrm{mag}(\omega)=\rho_0 v_s(\omega) \pi(\xi/2)^2,   
\end{equation}
where  $\rho_0$ is the density of scatterers, $v_s(\omega)$ is the frequency-dependent sound velocity. \textcolor{black}{We note that the simple spherical scattering cross section assumption only works well in the short wave length limit, i.e., the phonon wave length ($\lambda_p$) is smaller than $\xi$. This condition is valid in region II since $\xi>5$ $\mathring{\mathrm{A}}$ below 200 K as determined by inelastic neutron scattering experiments \cite{Williams2015}, whereas our calculations show that $\lambda_{p}\sim$ 2-5 $\mathring{\mathrm{A}}$. Normally, both $\rho_{0}$ and $\xi$ drops gradually with warming, and eventually become negligible above a characteristic temperature $T_0$ when short-range spin fluctuations diminish.} The exact forms of $\rho_{0}(T)$, $\xi(T)$ and $1/\tau_\mathrm{mag}(T)$ may depend on the details of magnetic interactions and dimensions, which are material-dependent. Here for CST, by fitting to the experimental $\kappa_(T)$ in region II, a simple form for $1/\tau_\mathrm{mag}(T)$ was found to be:
\begin{equation}
    1/\tau_\mathrm{mag}(T)=A(T_0-T)^\alpha v_s.
\end{equation}
 Here, $A$ and $\alpha$ are fitting constants, $T_0$ = 190 K is the temperature above which magnetic fluctuations become negligible, as extracted from the magnetic specific heat data. As seen in Figure~\ref{fig:4}c, phonon scattering rate due to spin fluctuations at 100 K is comparable to conventional scattering rate for phonon modes below 3 THz which dominate in $\kappa$. Consequently, spin-phonon scattering leads to a reduction of $\sim$ 50 \%  in $\kappa$. 

Now, we explore the field tuning effects. Figure~\ref{fig:4}d shows the calculated temperature- and field-dependent differential thermal conductivity ($\Delta\kappa$), which is defined as the difference between $\kappa_\mathrm{tot}$ evaluated with and without involving magnetic scattering, i.e. $\Delta\kappa=\kappa_\mathrm{tot}(1/\tau_\mathrm{c})-\kappa_\mathrm{tot}(1/\tau_\mathrm{c}+1/\tau_\mathrm{mag})$.  Clearly,  $\Delta\kappa$ weakens continuously with increasing magnetic field, both in regions I and II.  Since $\Delta\kappa$ always peaks around 20 K in all fields, we take the case at 20 K as an example to illustrate the magnetic field response of $\kappa$. \textcolor{black}{At 20 K, the system is in the FM ordered state, and the energy gap between the upper and lower magnon bands can be tuned by external magnetic fields. As a result, the resonant magnon-phonon scattering frequency $\omega_\mathrm{res}$ and scattering rate $1/\tau_\mathrm{mag}$ are field-dependent, leading to variable thermal conductivity in magnetic fields.} The energy gap in magnon dispersion increases gradually in finite magnetic fields, and a linear field dependence of $\hbar\omega_\mathrm{res}(B)$ is extracted by fitting the resonant scattering model to our experimental data (see Figure~\ref{fig:4}e). The fitted energy value $\hbar\omega_\mathrm{res}\sim$ 4 meV agrees quantitatively with that detected by neutron scattering studies \cite{Williams2015,Zhufengfeng2021}. In zero field, the absolute energy location at 8 meV of the lower band is inferred from Ref. \cite{Zhufengfeng2021}.   The linear field dependence of $\hbar\omega_\mathrm{res}(B)$ may be correlated with the topological nature of the magnon bands \cite{Zhufengfeng2021}. Similar effects have also been seen in few magnets carrying non-trivial magnetic excitations \cite{Hentrich2018,Yang2022}.   The resonant frequency $\omega_\mathrm{res}$ locates around 1 THz and shifts to larger values with increasing magnetic field (see the lower panel in Figure~\ref{fig:4}f). Importantly, as shown in the upper panel of Figure~\ref{fig:4}f, majority of heat is carried by phonons below 1 THz. Thus, as $\omega_\mathrm{res}$ moves above 1 THz, the destructive effects of magnon-phonon scattering on thermal transport is weakened. This leads to enhanced $\kappa_\mathrm{tot}(B)$ approaching gradually towards conventional phonon thermal conductivity $\kappa_\mathrm{tot}(1/\tau_\mathrm{c})$ in higher magnetic fields.  In region II, spins are partially polarized in magnetic fields, and short-range spin fluctuations are suppressed accordingly, as evidenced by substantial shifts of $g$ factor found by ferromagnetic resonance experiments \cite{Lizefang2021}. This naturally leads to reduced spin-phonon scattering strength, which again enhances $\kappa_\mathrm{tot}(B)$ in the presence of magnetic fields by partially recovering $\kappa_\mathrm{tot}(1/\tau_\mathrm{c})$.   

\begin{figure}[H]
\includegraphics[scale=0.25,center]{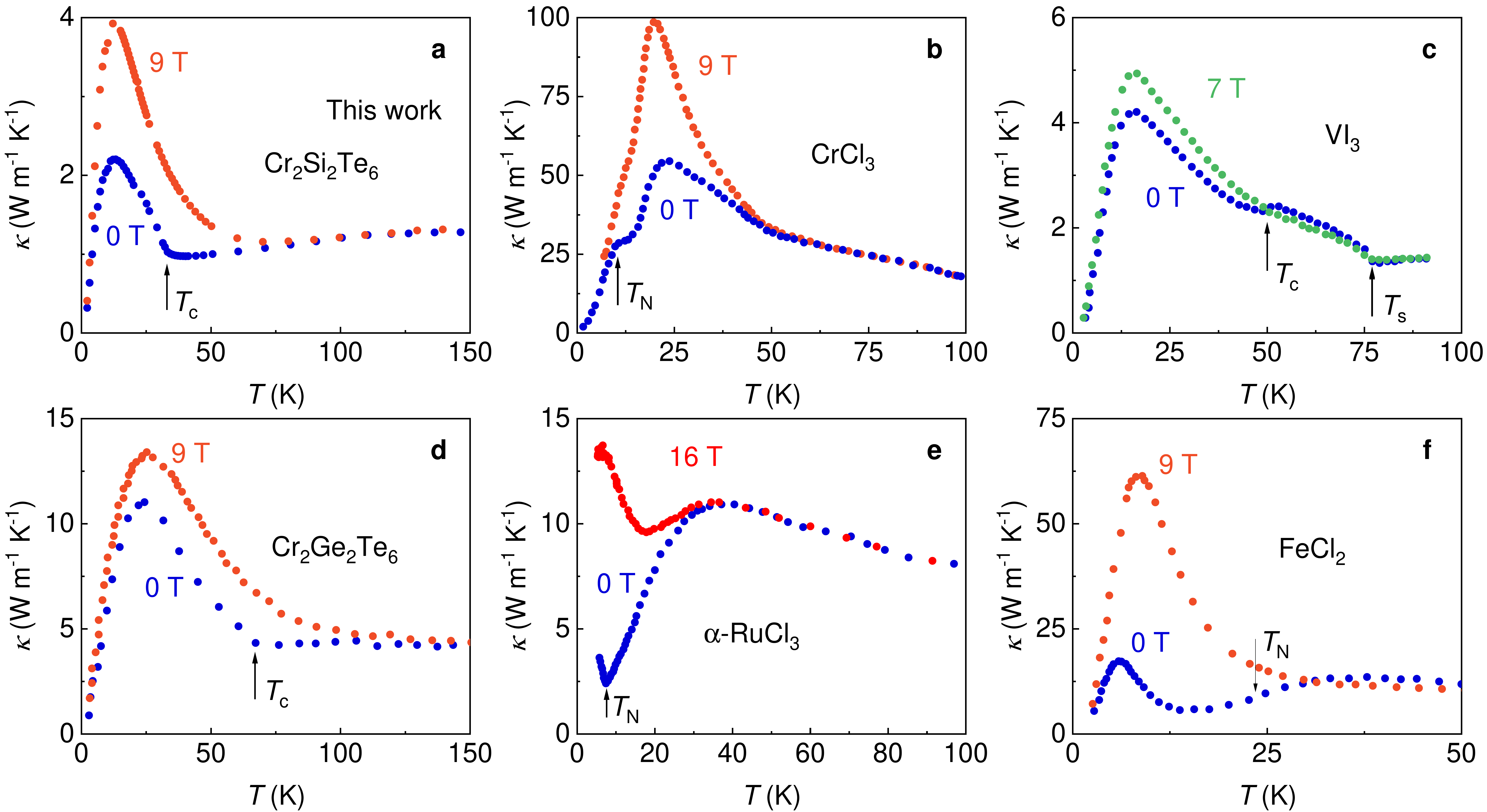}
\caption{\textbf{Comparison of thermal conductivity for typical vdW magnets.} Data shown for CrCl$_3$ \cite{Pocs2020}, VI$_3$ \cite{VI3_Thermal_Hall}, Cr$_2$Ge$_2$Te$_6$ \cite{Liuyu2022}, $\alpha$-RuCl$_3$ \cite{Hentrich2018} and  FeCl$_2$ \cite{FeCl2-Thermal-Hall} are from literature. The symbols $T_\mathrm{c}$, $T_\mathrm{N}$ and  $T_\mathrm{s}$ denote ferromagnetic, antiferromagnetic and structural transition temperatures, respectively.}
\label{fig:5}
\end{figure}

\begin{figure}[H]
\includegraphics[scale=0.5,center]{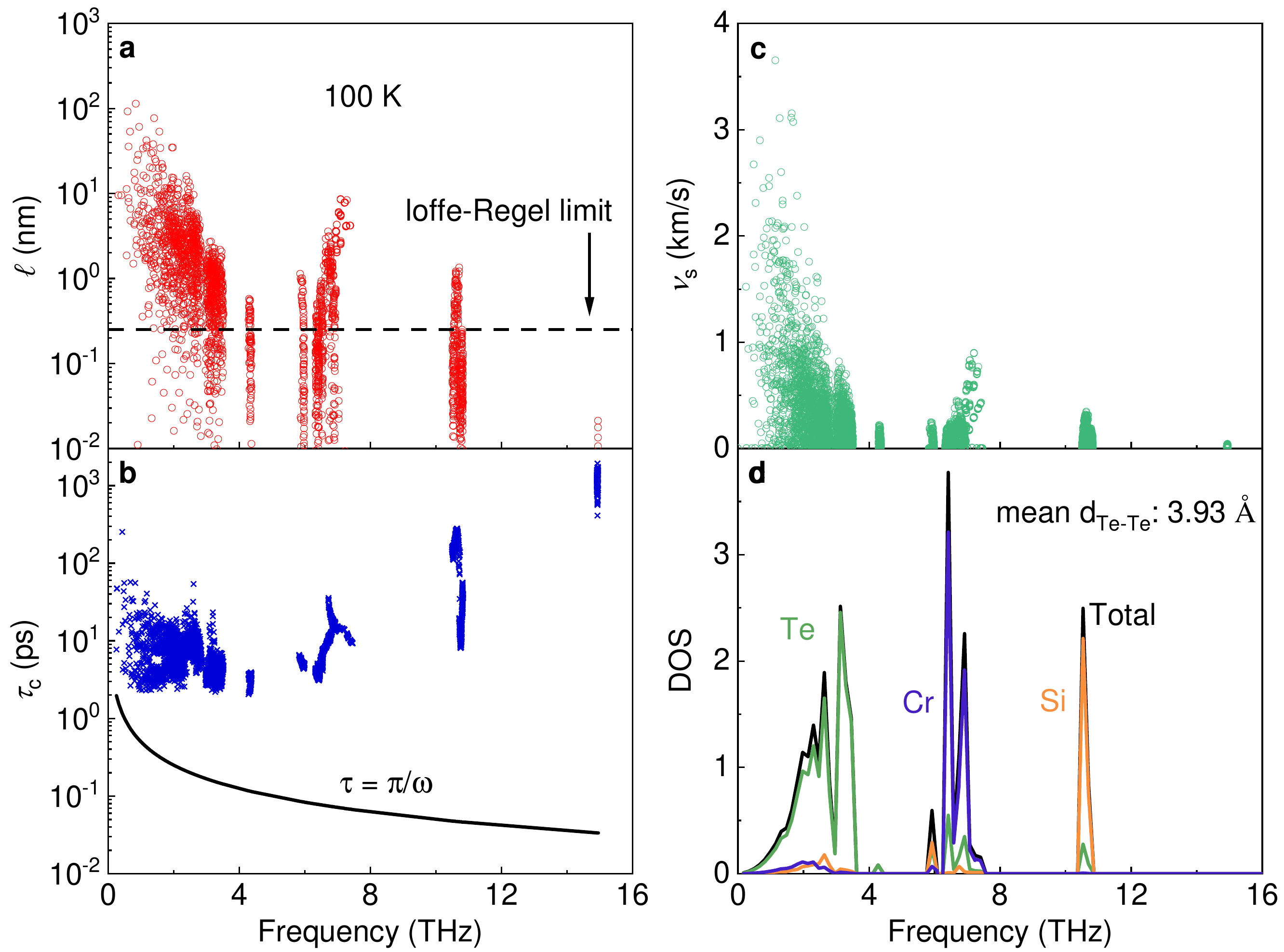}
\caption{\textbf{Intrinsically low phonon thermal conductivity of CST.} a) Phonon mean free paths obtained by DFT calculations. The black dash line marks the Ioffe-Regel limit. b) Frequency-dependent phonon lifetimes. The black solid line represents the minimum lifetimes. c) and d) Phonon group velocities and projected density of states (DOS). All data shown here were calculated at 100 K.}
\label{fig:6}
\end{figure}

\textcolor{red}{In Figure \ref{fig:5}, we compare the thermal conductivity of various typical vdW magnets. Note that these materials are insulating, and the thermal conductivity is dominated by phonon heat transport. Clearly, $\kappa$ is enhanced in the presence of magnetic fields for all these vdW magnets.  It is very likely that the same physics is at play, i.e., the spin-phonon scattering is suppressed by external magnetic fields, leading to field-induced enhancement of $\kappa$. These results imply that spin-phonon scattering is an effective channel in reducing phonon thermal conductivity, particularly in low-dimensional magnetic materials in which strong magnetic fluctuations naturally arise due to anisotropic magnetic interactions. Compared to other vdW magnets, the strength of spin-phonon scattering in CST may not be the strongest. However, the effects of spin-phonon scattering in CST survive at much higher temperatures up to about 200 K, leading to low thermal conductivity appearing within a wide temperature window. Similar effects are also seen in an isostructural sister compound Cr$_2$Ge$_2$Te$_6$ \cite{Liuyu2022} (Figure \ref{fig:5}d).  In CST, The magnetic $g$ factor is highly anisotropic and strongly temperature dependent \cite{Lizefang2021}. Compared with other vdW magnets, the temperature-dependent variations of the $g$ factor in CST are much larger and span over an exceptionally large temperature range \cite{Lizefang2021}. This may hold the key for the appearance of both static and dynamic spin fluctuations in CST even at room temperature \cite{Williams2015}.}

\textcolor{red}{In addition to spin-phonon scattering, the intrinsically low phonon thermal conductivity is another important factor guaranteeing the low thermal conductivity in CST over a wide temperature range. At high temperatures above 200 K, the spin-phonon scattering is negligible, and the measured thermal conductivity also stays around 1 W m$^{-1}$ K$^{-1}$ (see Figure \ref{fig:3}).  As seen in Figure \ref{fig:6}a, even at 100 K, the calculated phonon mean free paths (MFPs) $\ell$ without invoking magnetic scatterings are locating well below 100 nm. Considerable portions of phonon modes even show extremely low MFPs reaching the interatomic spacings (the Ioffe-Regel limit). As shown in Figure \ref{fig:6}b, the phonon lifetimes ($\tau_\mathrm{c}$) of all modes in CST are well above the minimum lifetimes defined by $\pi/\omega$ for highly disordered materials \cite{cahill1992}. However, the group velocities are rather low, showing an average value of about 1 km/s (see Figure \ref{fig:6}c). This indicates that the low phonon velocities are responsible for the short phonon MFPs and thus low thermal conductivity. From the projected phonon density of states (DOS) (Figure \ref{fig:6}d), it is seen that low-frequency heat-carrying phonons below 4 THz are mainly contributed by Te atoms. In CST, Te atoms are very weakly bonded, having a mean Te-Te spacing of 3.93 {\AA} \cite{CST-distance}, leading to low phonon velocities and flat phonon branches across the Brillouin zone (see Supplemental Material \cite{supplemental}). In the isostructural Cr$_2$Ge$_2$Te$_6$, the average Te-Te distance is 3.77 {\AA}  \cite{CGT-distance}, and Te atoms are linked stronger than those in CST. This likely accounts for the higher thermal conductivity of Cr$_2$Ge$_2$Te$_6$ than that of CST. Therefore, the intrinsically low phonon thermal conductivity, together with spin-phonon scattering, make CST a unique system for studying low thermal conductivity in vdW magnets.}

\section{Conclusion}
In conclusion, spin-phonon scattering-induced low thermal conductivity down to $\sim$1 W m$^{-1}$ K$^{-1}$ is achieved within a wide temperature range in Cr$_2$Si$_2$Te$_6$. The strength of spin-phonon scattering is tunable in the presence of an external magnetic field, giving rise to large thermal magnetoconductivity. Both the temperature dependence and magnetic field response of thermal conductivity are understood nicely by considering the temperature- and magnetic-field-dependent spin-phonon scattering effects. The superior low thermal conductivity and its easy tunability could further facilitate device applications of vdW magnets. Our theoretical understanding may also provide a general route for other systems in which spin-phonon scattering plays important roles.


\section{Methods}

\subsection{Sample growth}
Single crystalline Cr$_2$Si$_2$Te$_6$ samples were grown by the self-flux method with a molar ratio of Cr:Si:Te=1:2:6 \cite{Casto2015}. The mixture of pure chromium pieces (99.95\%, Kurt J. Lesker), silicon pieces (99.999\%, Kurt J. Lesker), and tellurium ingot (99.99\%, Alfa Aesar) were mounted in an alumina crucible, and further sealed inside a quartz tube under high vacuum ($<10^{-4}$ Pa). Then the starting materials were heated up to 1100 $^\circ \mathrm{C}$ in a shaft furnace and slowly cooled down to 700 $^\circ \mathrm{C}$. Excessive molten flux was centrifuged quickly before solidification.

 Bi$_2$Si$_2$Te$_6$ single crystals were grown by the self-flux method. Starting materials of Bi, Si, and Te were mixed in an Ar-filled glove box in a molar ratio of Bi:Si:Te = 1:1:13. The mixture  was  then loaded  in  an  alumina  crucible,  which  was  then  sealed in an evacuated quartz tube. The tube was heated up to 850 $^\circ \mathrm{C}$ in 10 hours and was kept there for 15 hours. Finally, the tube was slowly cooled down to 500 $^\circ \mathrm{C}$ with a cooling rate of 2 $^\circ \mathrm{C}$/hour, followed by  centrifuging to separate crystals from Te flux. Shiny and sizable crystals with hexagonal natural edges were eventually obtained.

 Crystals with typical dimensions of 3 $\times$ 1 $\times$ 0.2 mm$^3$ were chosen in the thermal transport, specific heat and electrical experiments.

 \subsection{X-ray diffraction}
 Single-crystal X-ray diffraction patterns were collected using a PANalytical XPert system working with Cu K$_{\alpha1}$/K$_{\alpha2}$ radiation ($\lambda_1 $ = 1.540598 $\mathrm{\mathring{A}}$ and $\lambda_2$ = 1.544426 $\mathrm{\mathring{A}}$).

 \subsection{Magnetization and specific heat measurements}
 Magnetization and specific heat experiments were performed in a Physical Property Measurement System (Quantum Design, Dynacool 9 T).  Magnetization was measured using the Vibrating Sample Magnetometer (VSM) option. Specific heat measurements were carried out using the standard relaxation method.

 \subsection{Thermal transport measurements}
 The thermal conductivity was measured using the steady-state technique on a homemade setup equipped with one heater and two thermometers.  \textcolor{red}{An optical image of the device is presented in Figure \ref{fig:3}d. The samples were glued to a copper heat sink using silver paint. Two thermometers were use to record the temperatures at the hot ($T_\mathrm{1}$) and cold ($T_\mathrm{2}$) ends. Samples and thermometers were thermally connected via silver wires (100 $\upmu$m in diameters) and silver paint.  } The thermometers have been carefully calibrated in the presence of magnetic fields. Low thermal conductivity PtW wires (30 $\upmu$m in diameters) were used for  electrical access to heater and thermometers. Temperatures and magnetic fields were controlled by a Physical Property Measurement System (Quantum Design, Dynacool 9 T). Similar thermal conductivity results were obtained using the Thermal Transport Option provided by the Physical Property Measurement System.

\subsection{Theoretical calculations}
The lattice thermal conductivity was calculated by using the state-of-the-art unified theory developed by M. Simoncelli \textit{et al}. \cite{Simoncelli2019}. Particle-like contributions from normal propagating phonons were obtained by using the exact solution to the Boltzmann transport equation (BTE) \cite{LI20141747,YANG2021100315,YANG2022100689,Yangxiaolong2021}. Scattering contributions from three-phonon, isotope, impurity, boundary,  magnon-phonon (below $T_\mathrm{c}$), and spin fluctuations (above $T_\mathrm{c}$) were considered in the calculations. Contributions from diffuson channels were computed by explicitly evaluating the off-diagonal terms of
the heat-flux operator \cite{Simoncelli2019}. Implementation of thermal conductivity calculations requires the harmonic and anharmonic interatomic force constants, which were calculated from density-functional theory within the generalized gradient approximation, using the first-principle-based
VASP package \cite{Kresse993}. See Supplemental Material for more details about theoretical methods, computational details, and determination of the adjustable parameters \cite{supplemental}.

\medskip
\textbf{Supporting Information} \par 
Supporting Information is available from the Wiley Online Library or from the author.

\medskip
\textbf{Acknowledgements} \par 
We thank Ping Li, Wenqing Zhang for stimulating discussions.  This work has been supported by National Natural Science Foundation of China (Grant Nos. 11904040, 52125103, 52071041, 12004254, 12004056, 11974065), National Key Research and Development Program of the Ministry of Science and Technology of China (Grant No. 2019YFA0704901), Chongqing Research Program of Basic Research and Frontier Technology, China (Grant No. cstc2020jcyj-msxmX0263), Chinesisch-Deutsche Mobilit\"atsprogamm of Chinesisch-Deutsche Zentrum f\"ur Wissenschaftsf\"orderung (Grant No. M-0496). 

\medskip
\textbf{Conflict of Interest}\par
The authors declare no conflict of interest.

\medskip
\textbf{Data Availability Statement}\par
The  data  that  support  the  findings  of  this  study  are  available  from  the  corresponding author upon reasonable request.

\medskip
\bibliographystyle{MSP}

\end{sloppypar}
\end{document}